\documentstyle[12pt]{article}

\begin{document}
\def\la{\mathrel{\mathpalette\fun <}}
\def\ga{\mathrel{\mathpalette\fun >}}
\def\fun#1#2{\lower3.6pt\vbox{\baselineskip0pt\lineskip.9pt
        \ialign{$\mathsurround=0pt#1\hfill##\hfil$\crcr#2\crcr\sim\crcr}}}
\newcommand {\eegg}{e^+e^-\gamma\gamma~+\not \! \!{E}_T}
\newcommand {\mumugg}{\mu^+\mu^-\gamma\gamma~+\not \! \!{E}_T}
\renewcommand{\thefootnote}{\fnsymbol{footnote}}
\bibliographystyle{unsrt}

\begin{flushright} \small{
UMD-PP-97-38\\
October, 1996}
\end{flushright}

\vspace{2mm}

\begin{center}
 
{\Large \bf Can Long Range Anomalous Neutrino\\
 Interactions Account for the Measured Tritium\\
 Beta Decay Spectrum ?}\\

\vspace{1.3cm}
{\large Rabindra N.\ Mohapatra$^1$    
and
Shmuel Nussinov$^2$}  

\vspace{5mm}
{\em $^1$ Department of Physics, University of Maryland, \\
College Park, Maryland 20742, USA\\ 
$^2$ School of Physics and Astronomy, Tel-Aviv University,\\
 Tel-Aviv, Israel\\}
\end{center}

\vspace{2mm}
\begin{abstract}

Recent experimental searches for neutrino mass in tritium beta decay
yield a negative value for the neutrino $(mass)^2$. If this effect
is genuine, then it is hard to understand it using conventional
particle physics ideas as embodied in the standard electroweak
model or its simple extensions that have been widely discussed.
We consider the possibility that there is a hidden anomalous
long range interaction of neutrinos that is responsible for this effect
and study the phenomenological consistency as well as tests of this idea.
We also discuss how such interactions may arise in extensions of the
standard model. 
 
\end{abstract}
 
\newpage
\renewcommand{\thefootnote}{\arabic{footnote})}
\setcounter{footnote}{0}
\addtocounter{page}{-1}
\baselineskip=18pt

\section{Introduction}

Several high precision experiments measuring the tritium end point spectrum
have been performed in order to put direct upper bounds on 
the mass of the electron neutrino, $m_{\nu_e}$. The
best fits of all experiments indicate however a negative mass squared for
the electron neutrino (i.e. $m^2_{\nu_e}<0$)\cite{rev}. Should this persist
in the future and manifest also in the end point spectra of other nuclei,
 an explanation
for this effect would clearly be called for. Needless to say that the
physics reasons for such an effect have to be very dramatic.

It has been pointed out by Stevenson et al\cite{stev} that the crossed process
in which a $\nu_e$ is absorbed from a background of electron neutrinos
\begin{eqnarray}
\nu_e + ^3H\rightarrow e^- + ^3 He
\end{eqnarray}
leads to electrons in the anomalous endpoint region. However in order to 
compete with the decay process $^3H\rightarrow ^3He + e^- +\bar{\nu_e}$
in this region, the required density of the background $\nu_e$'s should
be $n_{\nu_e}\simeq 10^{15}/cm^3$ or so i.e.
of order $|m_{\nu_e}|^3$ for $m_{\nu_e}\simeq 5$ eV, the
magnitude of the "imaginary-neutrino-best-fit-mass". 
This density vastly exceeds
that for gravitationally clustered neutrinos with (real) mass $m_{\nu_e}
\simeq 5$ eV\cite{gunn}. It could however be readily achieved if such
neutrinos would experience a local potential well
\begin{eqnarray}
U_{\nu_e}\simeq -|m_{\nu_e}|\simeq -(5-10)~eV
\end{eqnarray}
which fills up to a Fermi momentum $p_F\approx E_F\approx |U_{\nu}|$.

In the following, we will study various constraints on such new long
range forces involving neutrinos. We then discuss particle physics implications
and possible scenarios of new physics beyond
the standard model that may accomodate such unconventional interactions.

\section{Constraints on the new neutrino interactions}

The new neutrino interactions $U_{\nu}$ cannot exist uniformly
everywhere. A scalar uniform interaction simply
shifts ($m_{\nu}\rightarrow m_{\nu}+U_{\nu}$) the neutrino mass
whereas a uniform vector interaction would require
a background of almost conserved charge.

Independently of this, an enhanced cosmological neutrino density of such large 
magnitude as $n_{\nu_e}\approx 10^{15}/cm^3$ 
is completely at odds with the standard big bang scenario.
The measured $2.7^oK$ photon background implies
$n^{cosm}_{\nu}\approx 100/cm^3$. Thus the neutrino interaction energy
and the attendant enhancement of neutrino density 
can occur only in isolated regions
of typical size R, such that the total volume of these region comprises
a small fraction of the total volume now:
\begin{eqnarray}
\epsilon\leq {{n^{cosm}_\nu}\over{n^{local}_{\nu}}}\approx 10^{-13}-10^{-14}
\end{eqnarray}
The fraction of the Universe's volume occupied by galaxies is more than 
$\epsilon$ and therefore the possibility $R=R_{Galaxy}$ is ruled out
if all galaxies have the same enhanced density.
Furthermore the assumed $\nu_e$ mass of $5$ eV would lead to a dark
mass density $\rho_{\nu}\approx 5\times 10^{6}$ GeV or so which is
some seven orders of magnitude larger than the allowed value from observations
pertaining in particular to our galaxy.
However since the $\nu_e$ mass anomaly manifests in experiments at
vastly different locations on the Earth (Los-Alamos, Mainz etc), the
new interaction as well as the enhanced density should at least extend
over a region as large as the diameter of the Earth (i.e. $R\approx 10^9$ cm). 

The local neutrino interaction potential can be generated as a sum
of attractive pairwise potentials due to the exchange of a new, superlight,
boson of mass $\mu$. These exchanges can occur between the particular $\nu_e$
in question located, say, at ${\bf r}_{\nu_e}=0$ and other matter particles
(electrons, protons and neutrons) or with clustered neutrinos located at
$\bf {r\approx r_i}$:
\begin{eqnarray}
U^{\nu}_{local}({\bf{r}}=0)=\Sigma_i V(|\bf{r_i}|) 
\end{eqnarray}
The potential $V$ is taken to have the standard spin independent Yukawa 
form with a range $R\approx \mu^{-1}$ and strength factorizing to
the coupling to neutrinos $g_{\nu_e}$ and the coupling to matter particles
and neutrinos $g_{p,n,e,\nu_e}$. The exponential cut-off limits
the number of particles $(N_{p}, N_{n}, N_{e}, N_{\nu})$ 
 contributing to the above sum to those at locations
$|{\bf{r_i}}|\leq R\approx \mu^{-1}$. On general grounds of charge
neutrality, we have $N_p=N_e$. Also from stellar $n/p$ ratio (called 
$\zeta$), $\zeta\approx 1/7$ cosmologically and in the solar system
and  $\zeta\approx 1.2$ terrestrially,
we should have $\zeta\approx 0.14$ for $R\geq 1 A.U.\approx 1.5\times
10^{13}$ cm and $\zeta\approx 1.2$ for $1 AU\geq R\geq R_E\approx 10^{8}$ cm.
Under the natural assumption that the attracting particle distribution
extends at least up to the range of the force, the sum 
$U^{\nu}_{local} = \Sigma_i g_{\nu} g_i {{e^{-\mu r_i}}\over{r_i}}$
is dominated by $r_i\approx R\approx \mu^{-1}$. In order to generate
an $U^{\nu}_{local}\approx 5-10 eV$, we must therefore have,
\begin{eqnarray}
\mu g_{\nu}[N_p(g_e+g_p+g_n)+(N_{\nu}\pm N_{\bar{\nu}})]\geq 5-10 eV
\end{eqnarray}
For neutrinos we may have antiparticles contributing with same(opposite)
sign in the above equation depending on whether the interaction is
scalar or vector. 

If we assume that neutrino clustering is dominated
by the attraction of normal matter, it will naturally occur near the
Sun or the Earth as needed in order to explain the neutrino anomaly. The
condition in Eq.(5) can then be written as
\begin{eqnarray}
\mu N_p g_{\nu} g_m \geq 5-10 ~eV
\end{eqnarray}
where $g_m$ denotes the effective matter coupling. Clearly clustering will
be optimized if we use the the values of $R$ which maximize the number
$N_p$ of matter particles. One can contemplate two scenarios: (a) where
clustering arises from interactions on solar scale or (b) from
interactions on scale of the earth radius. In the
former case\footnote{We need not consider galactic scales for reasons stated
earlier. Precisely for this reason, we also did not consider possible
contribution to $U^{\nu}$ due to "wimps" or CDM particles etc.}, 
$R\approx 10^{13}$ cm and $N_p\approx 10^{57}$ whereas
in the latter case $R\approx 10^{9}$ cm and $N_p\approx 4\times
10^{51}$
From Eq. (6), we readily derive a lower bound on the strength of
the couplings $g_i$:
\begin{eqnarray}
g_m g_{\nu}\geq 0.8\times 10^{-38}~~~for~~ case (a)\\
g_m g_{\nu} \geq 0.5\times 10^{-36}~~~ for~~ case (b)
\end{eqnarray}

It is important to point out that alongside neutrino matter interaction,
the superlight boson exchange also mediates "diagonal" forces between ordinary
matter particles
\begin{eqnarray}
V_{mm}\approx g^2_m e^{-\mu r}/r
\end{eqnarray}
and between neutrinos
\begin{eqnarray}
V_{\nu\nu}= g^2_{\nu} e^{-\mu r}/r
\end{eqnarray}
On distance scales $r\approx R\approx 1~AU$ or $r\approx R\approx R_E$,
the $V_{mm}$ interactions can compete with the ordinary gravitational
interactions $V_{grav}\approx G_N m^2_p/r\approx 10^{-38}/r$. $V_{mm}$
can therefore spoil the equality of inertial and gravitational masses
which has been verified to an accuracy of one part in $10^{11}$ for
$R\approx 1$ A.U. (case (a)) and to one part in $10^{9}$ for $R\approx R_E$
(case (b))\cite{adel}. 
We can minimize the violation of the equivalence 
principle by artificially tuning 
$g_{i=e,p,n}$ to be proportional to the corresponding masses 
$m_{i=e,p,n}$. Even then the variation of nuclear binding energies
leads to a deviation from the equality of inertial and gravitational
masses at a level of $10^{-3}V_{mm}/V_{gr}$. From our previous
discussion, we then conclude that 
\begin{eqnarray}
g^2_m\leq 10^{-11+3}G_N m^2_p\leq 10^{-46}~~or~~ g_m\leq 10^{-23}
\end{eqnarray}
for case (a) and
\begin{eqnarray}
g^2_m\leq 10^{-9+3}G_N m^2_p\leq 10^{-44}~~or~~ g_m\leq 10^{-22}
\end{eqnarray}
for case (b).
Combining this with Eq. (7) and (8), we conclude that 
\begin{eqnarray}
g_{\nu}\geq 10^{-15}~~case(a)\\
g_{\nu}\geq 10^{-14}~~case(b)
\end{eqnarray}
These lead to rather "strong" long range "diagonal" $\nu\nu$ 
interactions. Such forces manifest in various settings:

\noindent{\bf (I):} For neutrinos of densities $n_{\nu}\approx p^3_F
\simeq U^3_{\nu}$ extending over large scales $R\approx \mu^{-1}$,
mutual $V_{\nu\nu}$ interactions could dominate over $V_{m\nu}$
and generate the requisite $U_{\nu}$
if 
\begin{eqnarray}
g^2_{\nu}N_{\nu}/R\approx g^2_{\nu}(RU_{\nu})^3/R\approx U_\nu\\ \nonumber
\end{eqnarray}
or
\begin{eqnarray}
g_{\nu}\approx {{1}\over{RU_{\nu}}}\approx \mu/ U_{\nu}\approx 10^{-18}/(R~in
AU)
\end{eqnarray}
However such self clustering of neutrino clouds could occur only for
scalar exchange interactions which generate attraction between $\nu\nu$
as well as $\nu\bar{\nu}$ pairs. The vectorial interactions are inherently
repulsive for any density of the relevant charges $\rho_{\nu}\equiv
n_{\nu}-n_{\bar{\nu}}$. Indeed one can easily show that 
\begin{eqnarray}
U_{vect}={{1}\over{2}}\int d^3r d^3r' \rho(\bf{r})\rho(\bf{r'})e^{-\mu(\bf
{r-r'}})/|\bf{r-r'}|\\ \nonumber
=\int d^3q \tilde{\rho}(\bf{q})\tilde{\rho}(\bf{q})(\mu^2+q^2)^{-1}\geq 0
\end{eqnarray}
where $\tilde{\rho}(q)$ and $(\mu^2+q^2)^{-1}$ are the Fourier transforms
of the neutrino density and the Yukawa potential.

\noindent{\bf (II):} Huge concentrations of neutrinos occur in the Supernovae
during gravitational collapse. Since all ranges R of the $V_{\nu\nu}$
and $V_{\nu m}$ considered 
exceed the supernova core radius ($R_{SN}\approx 10- 30~ Km$), 
we expect a mutual $\nu\nu$ interaction of order:
\begin{eqnarray}
W_{\nu\nu}\approx  N^2_{\nu}g^2_{\nu}/2R_{SN}
\end{eqnarray}
The neutrino densities and total numbers during the collapse are
comparable to those of nucleons. Yet this $W_{\nu\nu}$ should not
exceed the gravitational interaction during
the collapse 
\begin{eqnarray}
W_{grav}\simeq  G_Nm^2_pN^2_p/2R_{SN}
\end{eqnarray}
 in order not to disturb the standard supernova
dynamics which agrees pretty well with observations. Since $N_p\approx N_\nu$,
this would appear to lead to an independent bound on $g_\nu$
\begin{eqnarray}
g_{\nu}\leq (G_Nm^2_p)^{1/2}\approx 10^{-19}
\end{eqnarray}
At face value, this bound strongly conflicts with the minimal $g_{\nu}$
required, (see Eq.(13) and (14)). It turns out (as we show below)
that the bound in Eq.(19) holds only for vectorial interactions 
but not for scalar interactions.

\noindent{\bf Vector Interactions:}

\noindent For vectorial interactions, the
$N_{\nu}$ in Eq.(17) should be replaced by $\Delta N_{\nu}\equiv N_{\nu}-
N_{\bar{\nu}}$. The latter is roughly the total lepton number $N_L=N_e$
($=N_p$) of the collapsing core since a fair fraction of the $N_L$ is
trapped along with the $N_L=0$ thermally generated neutrinos in the core.
Thus it is clear that the above bound on $g_{\nu}$ applies in the case
of vectorial interactions.
[In passing we note that in the vectorial case, the "turning on" of the 
repulsive interaction upon core collapse is naturally avoided if the
almost massless vector boson couples to some conserved $U(1)$ charge.
The conservation of this new $U(1)$ charge for 
reactions such as $e^-+p\rightarrow
n+\nu_e$ responsible for neutrino production imply that the various $U(1)$
charges for the particles satisfy the relation\footnote{This is a consequence
of Weinberg's theorem extended to the case of $U(1)$ theories with almost
massless photons.}
\begin{eqnarray}
g_e+g_p=g_n+g_{\nu_e}
\end{eqnarray}
As a result, the total vector interaction energy is uneffected by
the reaction $e^-+p\rightarrow n+\nu_e$ throughout
supernova explosion process.
The condition that the neutron star not have substantial additional energy
due to $W_{nn}$ interaction still implies then that $g_n\leq 10^{-19}$. This
together with the constraint in Eq.(12) that $g_m\equiv g_p+g_n+g_e \leq
10^{-22}$ the condition in Eq.(20) implies that we cannot have
$g_{\nu}\geq 10^{-15}$ also in this case.]

\noindent {\bf Scalar Interaction:}

\noindent The situation is drastically different for scalar
interactions for several reasons: first, the scalar couplings need not 
satisfy any conservation laws like in Eq.(20). More importantly, the
estimate of the self interaction energy used above i.e. $W_{\nu\nu}
\approx g^2_\nu N^2_{\nu}/2 R$ is valid only for the case where
the neutrinos are mildly relativistic (i.e.$p_\nu\leq U_\nu\leq m_\nu$)
as in the putative neutrino cloud. However, for an extreme relativistic
neutrino gas as in the supernova core, the above expression for $W_{\nu\nu}$
is invalid. Let us consider a scalar exchange potential
between two neutrinos in the collapsing core. Because of their high energy
($E_{\nu_e}\geq 10 MeV\simeq 10^6 m_{\nu_e}$), the neutrinos are effectively
helicity eigenstates. The scalar exchange always flips helicity. Therefore 
to retain coherence implicit in adding all pairwise interactions, we need
to use the small $(m_{\nu_e}/E_{\nu_e})$ helicity admixture in
the wave function of the relativistic neutrinos. One therefore finds that
for relativistic neutrinos,
\begin{eqnarray}
W^{scalar}_{\nu\nu}\approx {{1}\over{2}}\left({{N^2_\nu g^2_\nu}\over{R}}
\right)\left({{m_\nu}\over{E_\nu}}\right)^2
\end{eqnarray}
In the supernova core, $(m_{\nu}/E_{\nu})^2\approx 10^{-12}$ making the upper
bound to $g_\nu\leq 10^{-13}$.
This bound is much more stringent than direct bounds on $g_{\nu_e}$
and $g_{\nu_{\mu}}$ implied by considerations of possible distortions
of the e spectrum in $\mu\rightarrow e\bar{\nu}_e \nu_{\mu}$\cite{pak};
 yet it allows for the 
anomalous long range interaction required for neutrino clustering. 
  
We thus conclude that it is phenomenologically allowed to have a scalar
interaction of neutrinos with strength $g_{\nu}\geq 10^{-15}$ that can
explain the apparent negative $(mass)^2$ puzzle of the neutrino experiments.
Let us therefore study possible particle physics implications of this idea.
 
\vspace{4mm}
\section{Particle Physics Implications:}
\vspace{4mm}
How likely is the possibility of such a scalar neutrino interaction from
particle physics point of view ? Since the force has a range of at least
$10^9$ cm, this implies that the mass of the scalar particle must be at
most $\mu\approx 10^{-14}$ eV (or for $R=1 AU$,
$\mu\approx 10^{-17}$ eV). Such small scalar masses are hard 
to understand since quantum corrections
often introduce infinite corrections to them 
thereby requiring extreme finetuning in each order of perturbation theory
to avoid large masses.
The second problem for the case at hand 
are the small scalar couplings which also
require a second fine tuning. To see the kind of fine tuning such small
values for $g_\nu$ would require in a generic $\lambda\phi^4$ theory, let
us note that we can retrieve the local potential due to a $\phi$ field as
\begin{eqnarray}
U_{\nu}\approx g_{\nu}\phi_{local}
\end{eqnarray}
Thus $\phi_{local}$ is given in the static approximation by 
\begin{eqnarray}
\phi_{local}=g_{\nu}\Sigma e^{-\mu r_i}/r_i
\end{eqnarray}
The positive energy density in the $\phi$ field given by $\lambda\phi^4$
 should not overwhelm the original negative energy $n_\nu U_\nu\approx
m^3_{\nu} U_{\nu}$ of the neutrinos. Using $U_\nu\approx m_{\nu}$, we
find that 
\begin{eqnarray}
\lambda \phi^4_{local}=\lambda (U_{\nu}/g_\nu)^4\leq
m^3_{\nu}U_{\nu}\approx U^4_{\nu} 
\end{eqnarray}
or finally\footnote{It is amusing to note that $\lambda\simeq g^4_{\nu}$
is precisely the self coupling induced by box diagrams with four external
$\phi$'s and four $\nu$ internal lines},
\begin{eqnarray}
\lambda\leq g^4_{\nu}
\end{eqnarray}
Even for $g_\nu$ saturating the supernova bounds, a very strong upper limit
of $\lambda\leq 10^{-52}$ is implied. Clearly it calls for an extreme degree
of fine tuning.

There are however field theories where these constraints on the
masses and coupling constants may be met in a natural manner.
We consider two examples below. In both cases,
 the scalar field is a pseudo-Goldstone boson which acquires
scalar couplings as well as a mass due to the presence of CP-violation.
The first example is a model proposed in Ref.\cite{CMN}, where it was
shown that the specific Goldstone boson,the singlet
Majoron\cite{cmp} which results from spontaneous
breaking of global $B-L$ symmetry can in the presence of the QCD anomaly,
acquire a mass. The majoron appears to have the right properties required for
our purpose. The second example is in the context of a general class
of pseudo-Goldstone models discussed by Hill and Ross\cite{hill}.

As is well-known, the Goldstone theorem requires that
for a theory with the Nambu-Goldstone boson, $\phi$
 the Lagrangian must be invariant under the transformation $\phi\rightarrow
\phi + \alpha $ where $\alpha$ is a constant. This implies that
the $\phi$ field must have zero mass and $\lambda=0$. Unfortunately, the
same invariance requirement also implies that the coupling of the
$\phi$ field is derivative type so that it eliminates the possibility
of having coherent $1/r$ type forces\cite{nus}.
 It actually leads to spin-dependent
forces\cite{cmp} in the non-relativistic limit. However, explicit
symmetry breaking via QCD anomalies\cite{CMN} 
not only generate small masses to
make the force finite range but also induce spin-independent couplings. 
In particular, in the model of Ref.\cite{CMN}, it was noted that the
scalar coupling of the Majoron to quarks is given by $g_Q\approx \theta
m_u/F_{\phi}\simeq 10^{-13}/(v_{B-L}~in~GeV)$, where $v_{B-L}$ is 
the $B-L$ symmetry
breaking scale. The mass of the majoron is given by $m_{\phi}\approx
g_Q \Lambda_{QCD}$. We note that if we choose $v_{B-L}\approx 10^{9}$ GeV, we
satisfy the bound on $g_m$ derived in Sec.II and a mass $m_{\phi}$ of the
right order (i.e. $m_{\phi}\approx 10^{-13}~eV$ that can lead to forces
with range $R\approx R_E$) 
is obtained. 

Let us now consider the coupling of the majoron
(now massive due to QCD anomalies) to neutrinos.
These couplings are given by $g_\nu\approx (m_{\nu_e}/ M_N)$ where $M_N$ is the
mass of the heavy right-handed neutrino. Choosing $m_{\nu_e}\approx 10$ eV
a right-handed neutrino mass of $10^6$ or $10^7$ GeV is required to obtain
$g_\nu\approx 10^{-14}$ or $10^{-15}$ as 
required to facilitate neutrino 
clustering. Since $M_N$ is related to $v_{B-L}$, it is
interesting that they are numerically not too far from each other.
In fact, if we choose a value for $\theta$ of about $10^{-12}$
 (instead of its maximum value $\theta\leq 10^{-10}$ 
allowed by present neutron electric dipole moment searches)
we would get $M_N=v_{B-L}$ making the model quite natural.

The second model is essentially an effective Lagrangian framework where one
uses the Goldstone boson corresponding to the spontaneous breaking of
of a chiral symmetry in conjunction with mass term for the Goldstone boson
and CP violation to generate the long range force. The basic idea
of the model can be demonstrated using
left and right-handed neutrinos and the chiral
lepton number symmetry as the broken symmetry. The effective Lagrangian
can then be written as
\begin{eqnarray}
L = L_0 + L_1 \\ \nonumber
\end{eqnarray}
where
\begin{eqnarray}
 L_0 = \bar{\nu}i\gamma^{\mu}\partial_\mu \nu + (m\bar{\nu}_L\nu_R e^{i\phi/F}
+h.c.) +1/2(\partial\phi)^2\\ \nonumber
\end{eqnarray}
and
\begin{eqnarray}
L_1 = \epsilon\bar{\nu}_L\nu_R+h.c. +\mu^2F^2 cos(\phi/F-\beta)
\end{eqnarray}
The effect of the $\mu^2$ term is to force $\phi$ to have a vacuum expectation
value. Defining $\tilde{\phi}=\phi-\beta F$ such that $\tilde{\phi}$ has
zero vev, one can rewrite the Lagrangian in terms of $\tilde{\phi}$.
The resulting Lagrangian has
a scalar coupling of $\tilde{\phi}$ to the neutrinos with strength
$g_\nu=\epsilon\beta/F$ and a mass for $\tilde{\phi}$ of $\mu$ which is an
arbitrary parameter. One can then choose the parameters $\epsilon, \beta$
and $F$ so as to get $g_{\nu}\approx 10^{-14}$. 
                                                
Let us now present a realization of this idea in a realistic extension
of the standard model. For simplicity let us only work with one 
generation and extend the standard model by adding a right-handed neutrino,
$\nu_R$ and a heavy neutral leptons $N_{L,R}$ as well as
a complex scalar boson $\Delta$ which is a singlet under the standard model
gauge group. Let us assume that the model has a global $U(1)$ symmetry
under which $\nu_R$ and $N_L$ have charges $+1$ and $-1$ respectively
and $\Delta$ has charge $+1$. The rest of the fields are neutral under it.
The Yukawa Lagrangian of this sub-sector of the theory is chosen to be
\begin{eqnarray}
L(\nu,N,\phi, H)= h\bar{\psi}_L H N_R +\bar{\nu}_RN_L\Delta^2/M 
+f\bar{N}_R\phi N_L\Delta + h.c.
\end{eqnarray}
where $\psi$ denotes the lepton doublet $(\nu,e^-)$ and $H$ is the
Higgs doublet of the standard model. The mass $M$ correspond to unknown
physics at a higher scale and is an unknown parameter for our model.
It is clear that after the electroweak symmetry
breaking and breaking of $U(1)$ symmetry by the vev $<\Delta>=F$ 
the $\nu$ and $N$ mix with each other. Writing the field
$\Delta={{1}\over{\sqrt{2}}}(F+\rho)e^{i\phi/F}$, we can obtain the
coupling of the physical neutrino fields with the Goldstone boson $\phi$
as follows:
\begin{eqnarray}
L_{\nu\nu\phi}\simeq m\bar{\nu}_R\nu_Le^{i\phi/F}
\end{eqnarray}
where $m\simeq {{F^3}\over{fM^2}}$. As in Hill and Ross\cite{hill}, let
us add to this theory the soft breaking terms in Eq.(26) which leads to the
desired long range forces. The important point here is that due to the
choice of our model, the soft breaking terms are all standard model singlets
and therefore do not spoil the successes of the standard model. Moreover
since the quarks or charged leptons
do not connect to the field $\Delta$, the light scalar has no coupling to
quarks or charged leptons.

\section{Conclusion and comments:}

In conclusion, scalar long range interactions of neutrinos required to
generate $U_\nu\sim 5-10$ eV and $n_\nu\simeq 10^{15}$ or $10^{16}~cm^{-3}$
on the solar system or Earth scale are not excluded by particle
physics considerations.
The bounds derived in this paper imply that neutrino self clustering
will generally dominate over clustering due to attraction of normal particles
in the Earth or the Sun. It is difficult to envision scenarios of capturing
such neutrino clouds onto the Earth (or the solar system). However in a recent
paper Stevenson et al\cite{goldman}\footnote{We thank V. L. Teplitz for
bringing this paper to our attention after our work was completed.} have
pointed out that a "role reversal" can occur according to which primordial
neutrino clouds can form first, once the temperature of the Universe drops below
$m_{\nu}$, before the baryonic matter can cluster. These clouds can then 
act as nucleation sites for the solar system. This interesting idea
deserves further investigation. It is however important to make the
following point in this connection: it is generally believed that our
solar system formed from a baryonic protocloud,
larger by about a factor 100 than the
present solar system. To efficiently assist in forming this protocloud, the
"seed neutrino cloud" would have to be about this size. If its density is
in the range considered above $\rho_\nu\simeq n_\nu m_{\nu}\approx
10^{16}-10^{17}$ eV/$cm^3$, then the total mass of the neutrino cloud would
range over $M_{\nu}\simeq (0.02-300) M_{\odot}$. Only a tiny portion of
the neutrino cloud mass could lie within the solar system ($R\simeq
3.10^{14}$cm) or at the inner planets ($R\leq R_E\simeq 1$ AU). These
values of extra dark mass (about $0.2 M_{Earth}-3.10^{-4}M_{Earth}$ do not yet
conflict with the recent Pioneer measurements and with precise orbit 
parameters for the inner planets found by radar ranging\cite{teplitz}.
However the effective total stellar masses seen by other stars would
include the full mass of the neutrino cloud. Studies of star clusters
could therefore exclude having $M_{\nu}\geq M_\odot$.

{\bf Acknowledgement}
\vspace{4mm}

The work of R. N. M. is supported by a grant from the National Science 
Foundation and the work of both R. N. M. and S. N. was also supported
in part by a grant from the US-Israel Bi-national
Science Foundation. 



\begin{thebibliography}{99}

\bibitem{rev} See talks by J. Bonn, V. M. Lobashev
and A. Swift in the proceedings of Neutrino'96, to be published by
World Scientific.

\bibitem{stev} R. G. H. Robertson, T. J. Bowles, G. J. Stevenson Jr.,
D. L. Wark, J. F. Wilkerson and D. A. Knapp, Phys. Rev. Lett. {\bf 67},
957 (1991).

\bibitem{gunn} S. Tremaine and J. E. Gunn, Phys. Rev. Lett. {\bf 42},
407 (1979)

\bibitem{adel} See E. Adelberger et al. Ann. Rev. Nucl. Part. Sc. {\bf 41},
269 (1991) for a review and earlier references.

\bibitem{pak} V. Barger, W. Y. Keung and S. Pakvasa, Phys. Rev. {\bf D25},
907 (1982); G. Gelmini, S. Nussinov and M. Roncadelli, Nucl. Phys. {\bf B209},
 157 (1982).

\bibitem{CMN} D. Chang, R. N. Mohapatra and S. Nussinov, Phys. Rev. Lett.
{\bf 55}, 2835 (1985).

\bibitem{cmp} Y. Chikashige, R. N. Mohapatra and R. D. Peccei, Phys. Lett.
{\bf 98B}, 265 (1981).

\bibitem{hill} C. Hill and G. G. Ross, Nucl. Phys. {\bf B311}, 253 (1988).

\bibitem{nus}  G. Gelmini, S. Nussinov and T. Yanagida, Nucl. Phys. {\bf B219},
31 (1983). 

\bibitem{goldman}  G. Stevenson Jr, T. Goldman and B. H. J. Mckeller,
hep-ph/9603392.

\bibitem{teplitz} V. L. Teplitz, private communication

\end{thebibliography}
\end{document}